\documentclass[pra,reprint,aps,twocolumn,nopacs,superscriptaddress,notitlepage]{revtex4-2}
\usepackage[utf8]{inputenc}
\usepackage[american,]{babel}
\usepackage[T1]{fontenc}
\usepackage[pdftex]{graphicx}  
\usepackage{graphicx, xcolor}
\usepackage{dcolumn}
\usepackage{bm}
\usepackage{amsmath,amsthm,amssymb}
\usepackage{hyperref}
\usepackage[T1,T2A]{fontenc}
\usepackage{xcolor}
\hypersetup{colorlinks,bookmarksopen,bookmarksnumbered,
    citecolor=blue,
    linkcolor=blue,
    pdfstartview=false,
    urlcolor=blue}
\usepackage{graphicx}
\usepackage{braket}
\usepackage{wrapfig}
\usepackage{soul}
\usepackage{mathtools}
\usepackage{float}
\usepackage{tabularx}

\renewcommand{\vec}{\mathbf}

\begin{document}
\newcommand{\titleinfo}{Entanglement and absorbing state transitions in $(d+1)$-dimensional stabilizer circuits} 
\title{\titleinfo}

\author{Piotr Sierant}
\affiliation{ICFO-Institut de Ci\`encies Fot\`oniques, The Barcelona Institute of Science and Technology, Av. Carl Friedrich Gauss 3, 08860 Castelldefels (Barcelona), Spain}
\author{Xhek Turkeshi}
\affiliation{Institut f\"ur Theoretische Physik, Universit\"at zu K\"oln, Z\"ulpicher Strasse 77, 50937 K\"oln, Germany}
\affiliation{JEIP, USR 3573 CNRS, Coll\`ege de France, PSL Research University,
11 Place Marcelin Berthelot, 75321 Paris Cedex 05, France}

\date{\today}
\begin{abstract}
We study the influence of feedback operations on the dynamics of $(d+1)$-dimensional monitored random quantum circuit. Competition between unitary dynamics and measurements leads to an entanglement phase transition, while the feedback steers the dynamics towards an absorbing state, yielding an absorbing state phase transition.
Building on previous results in one spatial dimension [\href{https://doi.org/10.1103/PhysRevLett.130.120402}{Phys. Rev. Lett. 130, 120402 (2023)}], we discuss the interplay between the two types of transitions for $d \ge 2$ in the presence of (i) short-range feedback operations or (ii) additional global control operations. 
In both cases, the absorbing state transition belongs to the $d$-dimensional directed percolation universality class. In contrast, the entanglement transition depends on the feedback operation type and reveals the dynamics' inequivalent features. The entanglement and absorbing state phase transition remain separated for short-range feedback operations. When global control operations are applied, we find the two critical points coinciding; nevertheless, the universality class may still differ, depending on the choice of the control operation. 
\end{abstract}

\maketitle

\section{Introduction}
Monitored many-body quantum systems provide a natural perspective for understanding the progress in quantum simulations~\cite{Fraxanet22} and noisy intermediate scale quantum technologies~\cite{preskill2018quantumcomputingin,ferris2022quantumsimulationon}.
Repeated measurements introduce non-unitary effects on the otherwise unitary evolution of quantum systems, leading to dynamics that can be described by stochastic quantum trajectories~\cite{carmichael2009open, Dalibard92,Molmer93, Wiseman2009,breuer2002}.
Crucially, there is a striking distinction between the average and typical properties of the trajectory ensemble. 
While the former lead to quantum channels and Lindbladian evolutions, the latter reveal a rich structure, including fingerprint phenomena like measurement-induced transitions (MIT)~\cite{cao2019entanglementina,skinner2019measurementinducedphase,li2018quantumzenoeffect,li2019measurementdrivenentanglement,chan2019unitaryprojective,sierant2022universalbehaviorbeyond}. 

The distinction between average and typical trajectory is of central importance for the observability of these transitions. 
While the average dynamics is experimentally feasible, extraction of the typical features of quantum trajectories requires post-selecting over the measurement results, a task of outstanding difficulty for generic systems and observables~\cite{koh2022experimentalrealizationof,fisher2022randomquantumcircuits,potter2022entanglementdynamicsin,lunt2022quantumsimulationusing,vasseur2019entanglementtransitionsfrom,jian2020measurementinducedcriticality,nahum2921measurementandentanglement,bao2020theoryofthe,gullans2020dynamicalpurificationphase,garratt2023probing}. 
Indeed, to perform the post-selection for a given quantum trajectory, one has to ensure that each of the conducted measurements yields the desired result. The MIT is observed in settings where the number of measurements scales proportionally to the space-time volume of the considered system. Since quantum measurements are inherently stochastic, the probability of obtaining a given trajectory is exponentially suppressed, or, in other words, the resources needed to perform an experiment scale exponentially with the system size.
Thus, without fine-tuning (cf.~\cite{sam.13.021026,hoke2023quantum,noel2022measurementinducedquantum,ippoliti2021postselectionfreeentanglement,passarelli2023postselectionfree}), avoiding or mitigating the post-selection is a central open problem in monitored quantum dynamics.

Recently, it has been proposed to use feedback operations that condition the system's dynamics on measurement outcomes to circumvent this post-selection problem. 
Indeed, conditional operations alter the average dynamics~\cite{friedman2023measurementinduced,roy1,roy2}, and, in principle, can encode non-linear features of the quantum trajectories, such as the MIT, even at the averaged density matrix level. 
This idea has been successful for monitored free fermions, and certain models of chaos~\cite{buchhold2022revealingmeasurementinduced,iadecola2022dynamicalentanglementtransition}, but the introduction of feedback does not necessarily imply that MIT is observable on the level of average state. 
For instance, when the feedback mechanism introduces an absorbing state to the system, i.e., a state that is a fixed point of the dynamics, the resulting absorbing phase transition (APT) and the MIT are generally distinct~\cite{odea2022entanglementandabsorbing,ravindranath2022entanglementsteeringin,ravindranath2023free,makki2023absorbing}. Nevertheless, for carefully chosen feedback operations \cite{sierantcontrol}, the fluctuations of the order parameter of the APT and the entanglement entropy can be coupled. In that case, the entanglement entropy undergoes a dynamical transition that inherits the universal features of the APT. However, even when the critical points of MIP and APT coincide, the universal content of the APT may differ from that of the MIT, depending on the renormalization group relevance of the underlying feedback operations~\cite{piroli,sierantcontrol}.

These previous works investigate one-dimensional systems and leave the role of dimensionality in monitored systems with feedback essentially unexplored.
Indeed, higher dimensional systems are generally challenging from a numerical perspective. The extensive entanglement generated by the weakly monitored dynamics poses severe limitations on tensor network methods~\cite{Schollw_ck_2011}. Similarly, the exponential growth of the Hilbert space with system size limits the exact simulations to a few tens of qubits. 
An important exception is stabilizer circuits that are efficiently simulable via the Gottesman-Knill theorem~\cite{aaronson2004improvedsimulationof,gidney2021stimfaststabilizer}, and have been recently investigated in $(d+1)$ random circuits with~\cite{turkeshi2020measurementinducedcriticality,sierant2022measurementinducedphase,lunt2021measurementinducedcriticality} and without monitoring~\cite{sierant2023entanglement}.

This paper investigates the interplay between the APT and MIT in $(d+1)$-dimensional stabilizer circuits. 
We employ the flagged Clifford circuits~\cite{piroli,sierantcontrol,ravindranath2023free}, showing that short-range feedback operations lead to distinct APT and MIT critical points and investigating their properties.
Subsequently, we also include a global feedback-control operation. In that case, the critical points of APT and MIT coincide. We unravel the similarities between the time evolution of the order parameter of APT and the entanglement entropy at any investigated dimension $d$. Finally, we discuss the range of validity of our results.

This manuscript is structured as follows. In Sec.~\ref{sec:2}, we review flagged stabilizer circuits, discuss the interplay between monitoring and feedback heuristically, and detail our implementation in $d\ge2$. The core section of our work is Sec.~\ref{supsec:3}, which discusses our numerical findings. 
Specifically, in Sec.~\ref{sec:3}, we study the order parameter behavior for the APT, highlighting its direct percolation (DP) universality class through numerical results in $2\le d\le 4$. In Sec.~\ref{sec:4}, we compare those findings with the entanglement dynamics for different choices of feedback operation. 
Sec.~\ref{sec:5} discusses the order parameter and entanglement entropy at a fixed circuit depth (time) $t \propto L$, revealing additional aspects of APT and MIT. Our conclusions, with further discussions and outlooks, are presented in Sec.~\ref{sec:6}.

\begin{figure}[t!]
    \centering
    \includegraphics[width=\columnwidth]{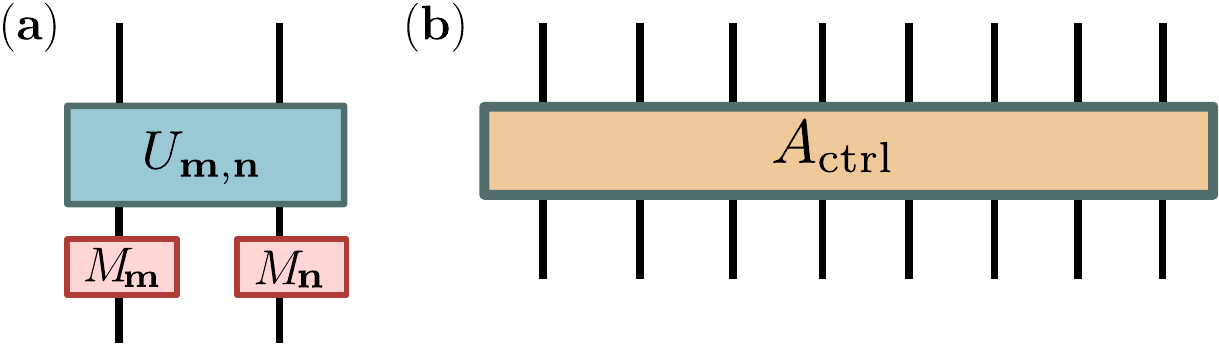}
    \caption{
    Gates building a layer $K=A_{\mathrm{ctrl}} K_0$ of the considered quantum circuit. The $K_0$ layer consists of gates $U_{\vec{m},\vec{n}}$ applied to neighboring qubits $\vec{m}$ and $\vec{n}$, depicted in panel (a). These gates include a two-body Clifford gate $U_{\vec{m},\vec{n}}$, conditioned on the flags $f_{\vec{m}}$ and $f_{\vec{n}}$ (as discussed in Sec.~\ref{subsec:flags}), as well as measurements $M_\vec{m}$ and $M_\vec{n}$ of the $Z_{\vec{m}}$ and $Z_{\vec{n}}$ operators. These measurements are performed with a probability of $p$. (b) Feedback-control operation $A_\mathrm{ctrl}$ is a global Clifford gate that acts non-trivially on all lattice sites at which $f_{\vec{m}}=0$.
    \label{fig:gates}
    }
\end{figure}

\section{Feedback-controlled and flagged stabilizer circuits} 
\label{sec:2}
This Section reviews the concept of flagged stabilizer circuits and details our numerical implementation of $d$-dimensional circuits. We also discuss here the phenomenology of our system.

\subsection{Flagged stabilizer circuits}
\label{subsec:flags}
We consider a $(d+1)$ dimensional quantum circuit defined on a $d$-dimensional spatial lattice $\Lambda$, comprised of $T$ layers that intersperse unitary dynamics and projective measurements of the local magnetization $Z_\vec{m}$. The lattice $\Lambda$ is fixed as as hyperrectangular $L_1\times\dots \times L_d$ lattice, where $L_1 = L$ and $L_{2}=\ldots=L_{d}=L/2$. We assume periodic boundary conditions in all directions and denote by $|X|$ the number of sites in a sublattice $X$. Throughout this manuscript, we denote the Pauli operators by $X_{\vec{m}}$, $Y_{\vec{m}}$,  $Z_{\vec{m}}$, while $\ket{1_\vec{m}}$ and $\ket{0_\vec{m}}$ are the +1 and -1 eigenvectors of $Z_{\vec{m}}$ and $\vec{m}$ labels the lattice sites.  
Each layer of the circuit is given by $K=A^{a}_{\mathrm{ctrl}} K_0$, where $K_0$ consists of the measurements and unitary gates, while $A_{\mathrm{ctrl}}$ is a feedback-control operation. We consider two cases: of short-range feedback, $a=0$ (and $A^{0}_{\mathrm{ctrl}}$ is an identity operator), and of a global control-feedback operation, $a=1$.

The measurement/unitary layer $K_0$ is built of $|\Lambda|/2$ two-body gates $U_{\vec{m},\vec{n}}$ presented in Fig.~\ref{fig:gates}(a). A two-body gate $U_{\vec{m},\vec{n}}$ acts on \emph{nearest neighboring} sites $\vec{m},\vec{n}$ of the lattice. The first index, $\vec{m}$, is chosen with uniform probability, without repetitions, over the whole lattice $\Lambda$. In contrast, the second index is set as $\vec{n} = \vec{m} +\vec{e}_u$ where $\vec{e}_u$ is a unit vector in a randomly chosen direction $u=1,\dots,d$. The gate $U_{\vec{m},\vec{n}}$ consists of the measurements $M_\vec{m}$, $M_\vec{n}$ of $Z_{\vec{m}}$, $Z_{\vec{n}}$ operators, and acts on system's state $\ket{\psi}$ via
\begin{equation}
    M_{\vec{m}}|\Psi\rangle  = \begin{cases}
        \displaystyle\frac{1+Z_\vec{m}}{2}\frac{|\Psi\rangle}{\sqrt{p_+}},\qquad \text{with probability }p_+\\
        \displaystyle\frac{1-Z_\vec{m}}{2}\frac{|\Psi\rangle}{\sqrt{p_-}},\qquad \text{with probability }p_-
    \end{cases}
\end{equation}
with ${p_\pm = \langle \Psi|1\pm Z_{\vec{m}}|\Psi\rangle/2}$ being the Born rule probability of the given measurement outcome. Each of the measurements is performed with the \textit{measurement probability} (or \textit{rate}) $p$, which is the control parameter that allows us to tune the considered quantum circuit between various dynamical phases.
The measurements are followed by an action of two-body gate $U_{\vec{m},\vec{n}}$ selected, with uniform probability, from the 2-qubit Clifford group. The gate $U_{\vec{m},\vec{n}}$ is conditioned on classical labels $f_{\vec{m}}$ in a way specified below. 
The procedure of formation of the layer $K_0$ consists of a random generation of $|\Lambda|/2$ two-body gates $U_{\vec{m},\vec{n}}$ that is performed independently during the construction of each of the layers of the circuit.

To introduce feedback in our system, we fix $|\mathrm{ABS}\rangle \equiv \bigotimes_{\vec{m}\in \Lambda} |1_{\vec{m}}\rangle$ as the absorbing state, i.e., we require that $|\mathrm{ABS}\rangle$ is a fixed point of the dynamics of our circuit, $K|\mathrm{ABS}\rangle = |\mathrm{ABS}\rangle$. 
For this purpose, the two-body gates $U_{\vec{m},\vec{n}}$ should preserve the $|1_{\vec{m}} 1_{\vec{n}}\rangle$ states. 
Since Clifford gates fulfilling this condition do not generate genuine quantum correlations, we introduce, following Ref.~\cite{sierantcontrol}, classical flags $f_{\vec{m}}=0,1$ at each site $\vec{m}\in \Lambda$ to establish the feedback mechanism in our stabilizer circuits.  
The system is initialized in the state $|\Psi_0\rangle = \bigotimes_{\vec{m}\in \Lambda}|0_{\vec{m}}\rangle $ and we initially set $f_{\vec{m}} = 0$ for all $\vec{m}\in \Lambda$. After each measurement, we change the flag to $f_{\vec{m}}=1$ when the outcome is $+1$ (otherwise, the flag remains unchanged, $f_{\vec{m}}=0$). 
The two-body gate $U_{\vec{m},\vec{n}}$
acts on the sites $\vec{m},\vec{n}$ only when $f_{\vec{m}}f_{\vec{n}}=0$. Otherwise, $U_{\vec{m},\vec{n}}$ is replaced by the two-site identity matrix.

A short-range feedback mechanism is present in a circuit comprised solely of layers $K_0$ due to the flag mechanism built in the two-body gates $U_{\vec{m},\vec{n}}$. It is straightforward to verify that $\ket{\mathrm{ABS}}$ is indeed an absorbing state: $K_0  \ket{\mathrm{ABS}}=\ket{\mathrm{ABS}}$.

Finally, to introduce a global control operation to our system, we consider $A_\mathrm{ctrl}$, see Fig.~\ref{fig:gates}(b), which is a global random Clifford unitary that acts non-trivially only on the subset $\tilde{\Lambda} = \{\vec{m}\in \Lambda: f_{\vec{m}}=0\}\subset \Lambda $ of unflagged sites. 
This construction of the feedback-control operation $A_\mathrm{ctrl}$ ensures that our stabilizer circuit can generate extensive entanglement in the presence of monitoring while preserving $|\mathrm{ABS}\rangle$ as the absorbing state. In the following, we will compare and contrast the properties of the circuit built of layers of $K=K_0$ with the time evolution of the circuit $K=A_\mathrm{ctrl } K_0$   composed of the global feedback-control operation $A_\mathrm{ctrl}$ and the measurement/unitary layer $K_0$.

We note that with these specifications, the described setups are amenable to efficient numerical simulations for $d\geq 2$~\cite{sierant2022measurementinducedphase,sierant2023entanglement} that scale polynomially in the system size $L$. 
Our simulations of flagged stabilizer circuits are implemented in a state-of-the-art package \textsc{STIM}~\cite{gidney2021stimfaststabilizer} and employ an 
asymptotically fast~\cite{andren2007onthecomplexity,albrecht2011efficient} algorithm for computation of rank with complexity $O(N^3 / \log_2 N)$~\cite{Bertolazzi14}, where $N=|\Lambda|$ is the number of qubits in the lattice.

\subsection{Post-selection: linear and non-linear functions of the density matrix}
\label{subsec:phenom}

Before dwelling into the systematic numerical analysis of the next Section, we would like to highlight some vital physical aspects of the dynamics of the considered quantum circuits with feedback. 
Performing numerical simulations of quantum dynamics of the flagged stabilizer circuits, we obtain the time-evolved state $\ket{\Psi_t} \equiv \prod_{t} K_t \ket{\Psi_0}$, and the corresponding density matrix $\rho_t \equiv |\Psi_t\rangle\langle\Psi_t|$.
We are interested in quantities that are averaged over the circuit realizations. This leads to the crucial difference between physical quantities concerning their dependence on the density matrix $\rho_t$:
\begin{itemize}
    \item linear functions of $\rho_t$, for instance a defect density 
    \begin{equation}
        n_\mathrm{def}(\Psi_t)=1- \mathrm{tr} \left( \rho_t  \sum_\vec{m} \frac{1+Z_ \vec{m}}{2N } \right),
        \label{eq:Def}
    \end{equation}
    where $N= |\Lambda|=L(L/2)^{d-1} $ is the total number of sites in the lattice.
    Taking the average (denoted by the overline) over the circuit realizations of the defect density yields $n_\mathrm{def} \equiv \overline{ n_\mathrm{def}(\Psi_t)}$, which due to the linearity of the considered quantity amounts to  
    \begin{equation}
        n_\mathrm{def}=1- \mathrm{tr} \left( \overline{ \rho_t }  \sum_\vec{m} \frac{1+Z_ \vec{m}}{2N } \right),
    \end{equation} 
    i.e., the average defect density $n_\mathrm{def}$ is determined solely by the average density matrix $\overline{ \rho_t }$.
    \item non-linear functions of $\rho_t$, for instance entanglement entropy
    \begin{equation}
        S_X(\Psi_t ) = -\overline{\mathrm{tr}_X(\rho_X(t)\log_2\rho_X(t))},
        \label{eq:Ent}
    \end{equation}
     where $\rho_X(t)=\mathrm{tr}_{X_c}(\rho_t)$ is the reduced density matrix for subsystem $X$~\cite{breuer2002} obtained by tracing out the degrees of freedom of its complement $X_c$ ($\Lambda = X \cup X_c$). Due to the non-linearity of \eqref{eq:Ent}, the average entanglement entropy, $S_X(t)\equiv \overline{ S_X(\Psi_t ) }$, has to be calculated directly by evaluation of $S_X(\Psi_t )$ and by averaging the result over the circuit realizations. In other words, there is generally no functional dependence between $S_X(t)$ and the average density matrix $\overline{ \rho_t }$.
\end{itemize}
The dichotomy between the linear and non-linear functions of $\rho_t$ is reflected on the level of physical quantities and phenomena that can be captured with the two types of quantities.
The averages of linear functions of  $\rho_t$ are amenable to experiments as they do not require the post-selection and are dependent solely on the average density matrix $\overline{ \rho_t }$. The defect density $n_{\mathrm{def}}$ captures the APT in the system.
Conversely, the non-linear functionals of the state, such as the entanglement entropy $S_X(t)$, reveal phenomena that occur on the level of individual trajectories of the system, such as the MIT, and require the post-selection. Indeed, to calculate $S_X(\Psi_t ) $, we have first to ensure that we are considering a fixed final state $\ket{\Psi_t}$ that depends on the outcomes of the performed measurements, then evaluate $S_X(\Psi_t )$ by repeatedly preparing the same final state $\ket{\Psi_t}$, and only then we can average the result over the circuit realizations.

\begin{figure}[t!]
    \centering
    \includegraphics[width=\columnwidth]{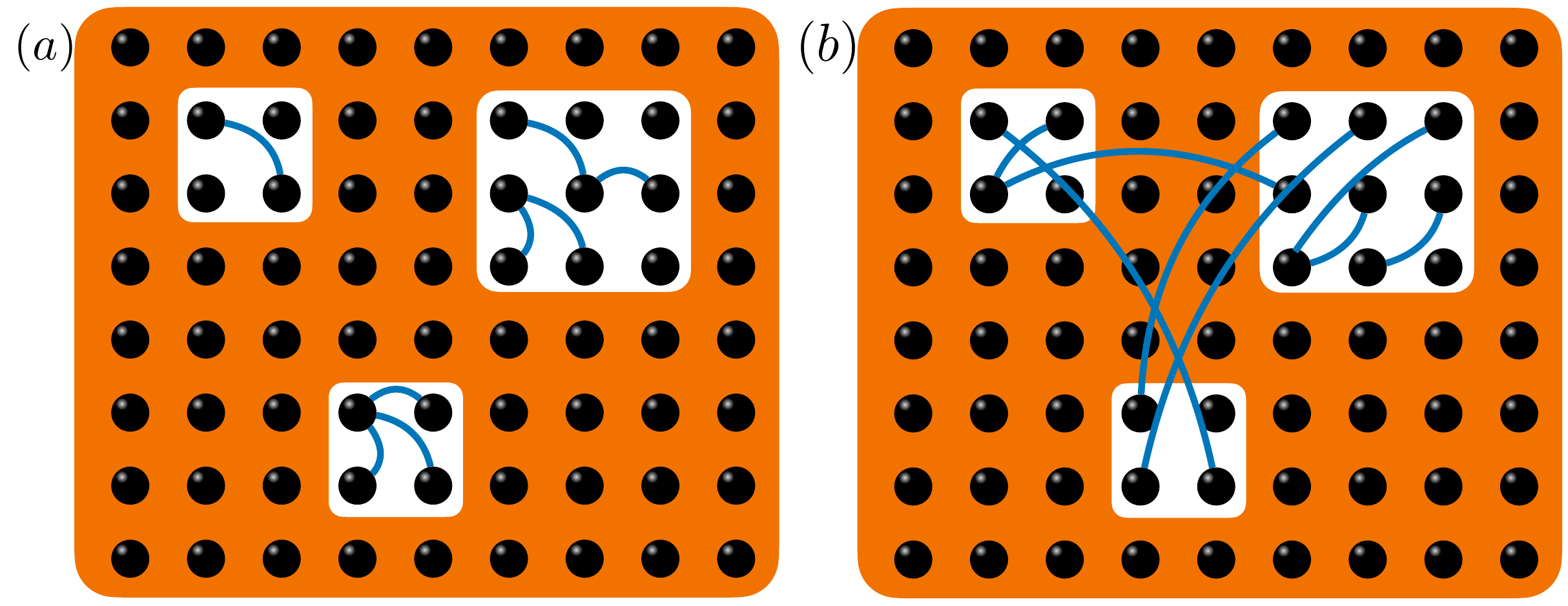}
    \caption{
     Conditioning of the unitary gates on the measurement outcomes by the flags mechanism leads to the emergence of ordered domains (highlighted in orange) and defect regions. (a) Short-range control operations only entangle degrees of freedom within the same defect regions. (b) Additional global feedback-control operation $A_\mathrm{ctrl}$ generates long-range entanglement, coupling distant disordered areas. Blue lines pictorially represent entangled degrees of freedom.
    \label{fig:cartoon_fig}
    }
\end{figure}
\subsection{Phenomenology of feedback-monitored systems with an absorbing state}

In our system, due to the presence of the feedback mechanism, we expect that $\lim_{t\to\infty} {\rho} = |\mathrm{ABS}\rangle\langle\mathrm{ABS}|$. Indeed, if in particular region $\lambda$ of the lattice $\Lambda$ the measurements of $Z_\vec{m}$ yield the result $1$, the flags in the region $\lambda$ are set to unity, $f_\vec{m}=1$. Hence, due to the employed feedback, the unitary gates $U_{\vec{m}, \vec{n} }$ can act non-trivially only at the edges of the region $\lambda$. In contrast, in the bulk of the subsystem $\lambda$, the state is already locally ordered ferromagnetically as in the absorbing $|\mathrm{ABS}\rangle$. Hence, over time evolution of our system, the lattice $\Lambda$ becomes covered with the ordered domains, in which the spins are aligned as in the absorbing state, $\langle \Psi_t|Z_{\vec{m}}|\Psi_t\rangle = 1$, and defect regions in which the spins are not aligned in that way, $\langle \Psi_t|Z_{\vec{m}}|\Psi_t\rangle \neq 1$, as schematically presented in Fig.~\ref{fig:cartoon_fig}. The fraction of sites in the defect regions is given precisely by $n_{\mathrm{def}}$ defined in~\eqref{eq:Def}.

Since $|\mathrm{ABS}\rangle$ is invariant under each layer $K$ of the circuit, the ordered domains, on average, grow. This introduces ordering to the system, which finally reaches the absorbing state. The timescale for reaching $|\mathrm{ABS}\rangle$ is altered by $p$.
At high measurement rates ($p>p_c^\mathrm{APT}$), the ordered regions develop quickly, and the defect density $n_{\mathrm{def}}$ decays exponentially in time. 
Conversely, at small measurement rates ($p<p_c^\mathrm{APT}$), the system is in a non-absorbing phase as the unitary gate scramble information while competing with the measurements and a non-vanishing defect density $n_{\mathrm{def}}$ prevails to time scales that grow exponentially with the system size.
Close to the critical point of APT $p \approx p_c^\mathrm{APT}$, the ordering in the system develops so that the defect density decays in a characteristic power-law fashion, which is a signature of the APT.

The dynamics of the APT can be observed on the level of the average state $\overline{\rho_t}$. In contrast, the entanglement properties of the system unravel a richer structure observable on the level of individual trajectories $\ket{\psi_t}$.

Notably, the entanglement content of the system is fixed by the presence/absence of the feedback-control operation $A_{\mathrm{ctrl}}$, which, by construction, does not affect the dynamics of the average state $\overline{\rho_t}$. 
Without the control operation ($a=0$, $K=K_0$), the feedback mechanism is solely short-ranged, and no-long range entanglement between distant disordered regions is generated, cf.~\ref{fig:cartoon_fig}(a). Indeed, the unitaries $U_{\vec{m}, \vec{n} }$ generate quantum correlations only among degrees of freedom within or close to the boundary of defect regions. 
The absorbing state is a product state. Hence, we expect that the MIT occurs before the APT (i.e., $p_c^\mathrm{MIT}<p_c^{\mathrm{APT}}$) in such a way that the state can follow the area-law of entanglement entropy while the system is not yet in an absorbing phase. For instance, the state may host isolated single-site defects. Such a state is not volume-law entangled but is still not an absorbing state. 

When the global feedback-control operation $A_\mathrm{ctrl}$ is employed, it couples globally all defect regions, creating long-range entanglement between distant defects cf. Fig.~\ref{fig:cartoon_fig}(b). In this case, we expect $p_c^\mathrm{MIT}=p_c^\mathrm{APT}$, since any arbitrarily separated qubits in a defect state will be correlated by $A_\mathrm{ctrl}$ and only a fully ordered state hosts no entanglement. This heuristic discussion was corroborated for $d=1$ dimensional systems in \cite{sierantcontrol}.
We confirm this picture with a systematic numerical analysis for $d \geq 1$ in the following sections.

\section{Numerical results}
\label{supsec:3}
In this Section, we discuss the numerical results for the described circuit architecture and various space dimensions $d \leq 4$, considering the average dynamics reflected by the density of defects $n_{\mathrm{def}}$ as well as the non-linear functions of $\rho_t$. 
Specifically, we investigate the entanglement entropy $S_X(t)$ dynamics for the setups with short-range feedback mechanisms and with the additional global feedback-control operation $A_\mathrm{ctrl}$. Lastly, we investigate the system's entanglement, and average state features at a fixed circuit depth.

\begin{figure*}
    \centering
    \includegraphics[width=1\linewidth]{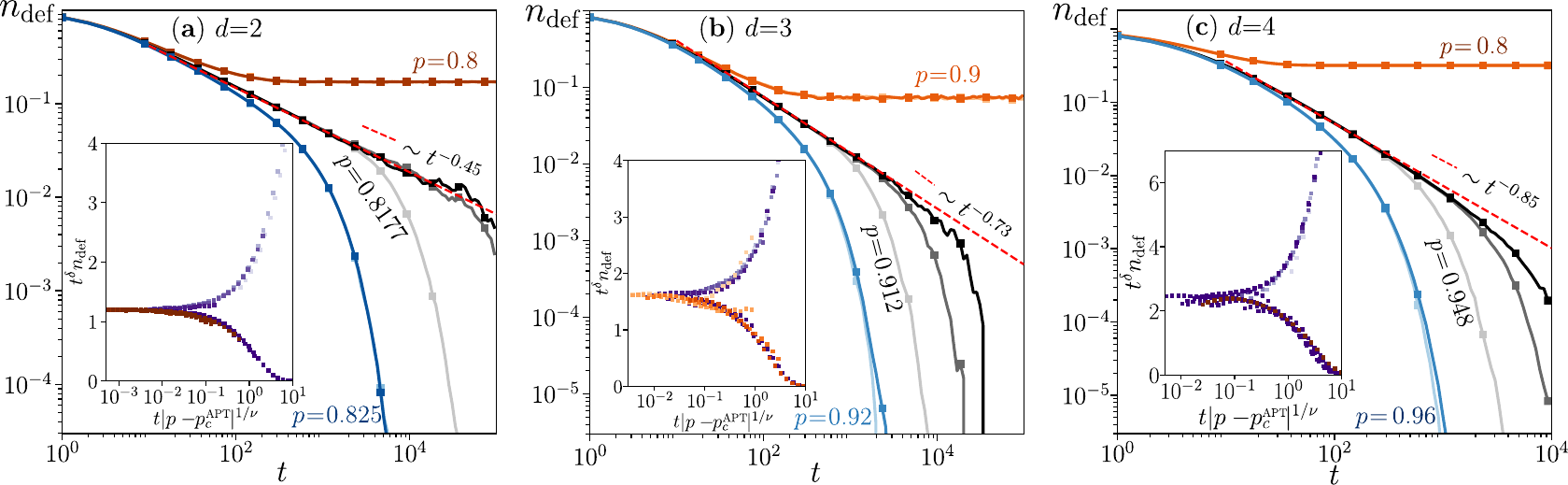}
    \caption{Absorbing state phase transition in $d=2,3,4$ dimensional circuits. The time evolution of the defect density $n_{\mathrm{def}}$ was obtained by considering the PCA for the average dynamics (see Text). The main panels show the $n_{\mathrm{def}}$ as a function of time $t$ for measurement rate below/close to/above the APT. The darker colors correspond to increasing system sizes.
    For $d=2$ (a) we consider $L=100,400,800$, for $d=3$ (b) $L=32,64,128$, and for $d=4$ (c) $L=16,32,64$. 
    In the vicinity of the APT, we observe the characteristic power-law decays $n_{\mathrm{def}} \propto t^{-\delta}$ with exponents $\delta$ close to the exponents for DP class. The insets show $t^{\delta} n_{\mathrm{def}}$ plotted as functions of $t|p-p^{\mathrm{APT}}_c|^{1/\nu}$, demonstrating data collapses with exponents consistent with the DP class, see Tab.~\ref{Tab1} for the values of the exponents.
    }
    \label{fig2}
\end{figure*}

\subsection{Dynamics of the order parameter}
\label{sec:3}
We begin by analyzing the dynamics of the average state $\overline \rho_t$. Following the standard techniques, see e.g.~\cite{odea2022entanglementandabsorbing,piroli,ravindranath2022entanglementsteeringin,sierantcontrol}, the average dynamics can be mapped analytically to a probabilistic cellular automaton. Appendix~\ref{app:mapping} details a short discussion about this mapping which we use to calculate quantities depending on $\overline \rho_t$, such as the defect density $n_\mathrm{def}$. The computation of the classical average dynamics is more efficient than calculating the full quantum dynamics of the circuit with stabilizer formalism, which allows us to simulate systems in $2\le d\le 4$. 
Focusing on the evolution of the defect density $n_\mathrm{def}$ with the time (circuit depth) $t$ close to the APT, 
we average the results over no less than $N_\mathrm{real}=200$ realizations of the circuit and study the behavior $n_\mathrm{def}$ in systems of size up to $L=800$ in $d=2$, $L=128$ in $d=3$, and $L=64$ in $d=4$ dimensions. Our results are summarized in Fig.~\ref{fig2}. 

Our results for $d=2$ and systems of size $L\geq 100$ are shown in Fig.~\ref{fig2}(a). At $p=0.8$, the defect density attains a non-zero stationary value that persists to time scales exponentially increasing with $L$, indicating that the system is in the non-absorbing phase. In stark contrast, a hallmark of the absorbing phase is visible for $p=0.825$: the defect density decays to zero exponentially with time $t$ independently of the system size $L$. The critical point that separates the two phases is located at a measurement rate $p_c^{\mathrm{APT}} = 0.8175(2)$ at which a power-law decay $n_\mathrm{def} \propto t^{-\delta}$ with an exponent $\delta=0.45(1)$ emerges. This behavior is characteristic for the DP universality class in $d=2$. Varying the mesurement rate around $p=p_c^{\mathrm{APT}}$, we observe a collapse of $t^{\delta} n_{\mathrm{def}}$ plotted as a function of $t|p-p_c^{\mathrm{APT}}|^{1/\nu}$, see the inset in Fig.~\ref{fig2}(a), with exponent $\nu=1.30(3)$, in agreement with the $d=2$ DP universality class~\cite{Munoz99, Hinrichsen00}.

Our results for the average dynamics in $d=3$ are presented in Fig.~\ref{fig2}(b). In the non-absorbing phase, the density of defects $n_{\mathrm{def}}$ attains a non-zero stationary value for up to a time scale exponentially large in $L$, as exemplified by the results displayed for $p=0.9$. In the absorbing phase, $n_{\mathrm{def}}$ decreases exponentially to zero, as demonstrated by the data for $p=0.92$. At the APT in $d=3$, at $p_c^{\mathrm{APT}}=0.912(1)$, we notice a power-law decay $n_{\mathrm{def}} \propto t^{-\delta}$, with the exponent $\delta = -0.73(2)$ compatible with the $d=3$ DP universality class. Moreover, as the inset in Fig.~\ref{fig2}(b) illustrates, we find a collapse of $t^{\delta} n_{\mathrm{def}}$ versus $t|p-p_c^{\mathrm{APT}}|^{1/\nu}$ with $\nu = 1.11(4)$ consistently, within the error bars, with the critical exponents for $d=3$ DP~\cite{Munoz99, Hinrichsen00}.

As shown in Fig.~\ref{fig2}(c),  the defect density $n_{\mathrm{def}}$ in $d=4$ dimensional system behaves in a quantitatively similar fashion in the non-absorbing phase (e.g., at $p=0.8$) and in the absorbing phase (e.g., at $p=0.96$). The two phases are separated by a phase transition at which a power-law decay of the defect density emerges. At the considered system size $L=64$, we find that the decay of $n_\mathrm{def}$ is well approximated by a decay with an exponent $\delta=0.85(1)$ at $p^{\mathrm{APT}}_c=0.948(2)$. However, by comparing this exponent with the results for $L=16$ and $L=32$, we notice a persistent increase of our estimate of $\delta$ with increasing $L$. For instance, at $L=16$, the power-law decay persists for the longest time for $p=0.945$ with exponent $\delta =0.79(1)$. Hence, the effects of finite system size introduce a systematic error into our numerical analysis, preventing us from quantitatively confirming the mean-field critical exponents $\delta=1$ and $\nu=1$ expected for DP universality class at $d=4$~\cite{henkel2008non}. Nevertheless, the trends characterizing our results suggest that the mean-field critical exponents may describe the considered system when the time scales and system sizes are sufficiently large.

\begin{figure*}[t!]
    \centering
    \includegraphics[width=1\linewidth]{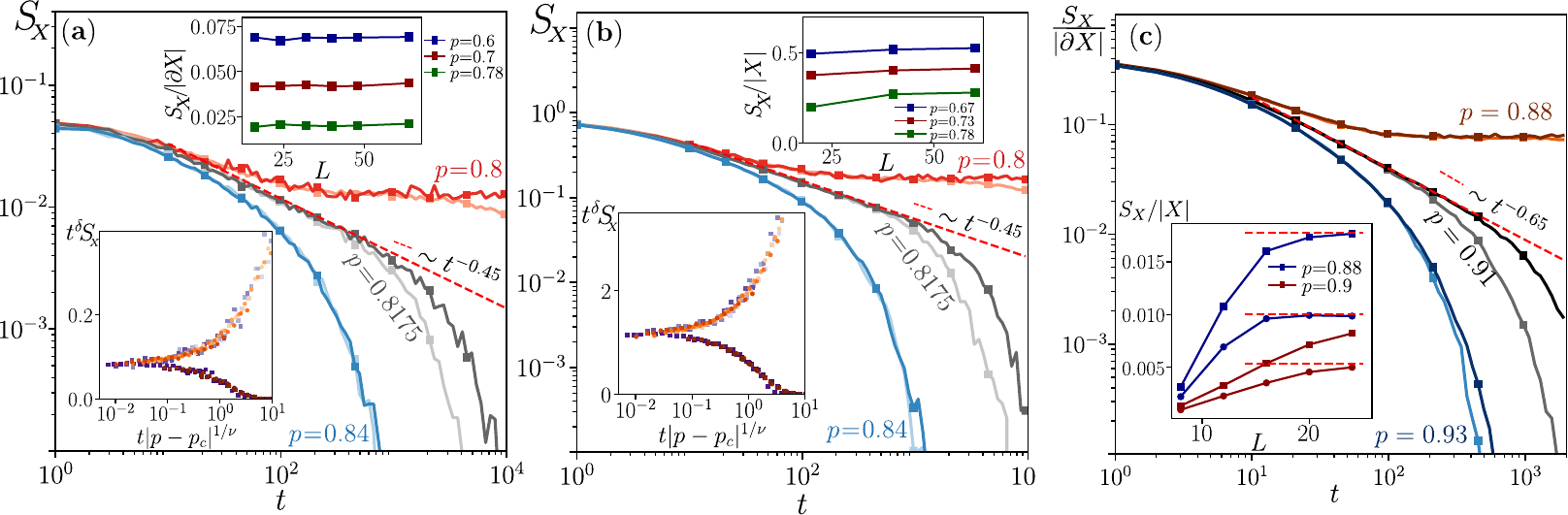}
    \caption{Entanglement entropy $S_X$ without the control operation  (\text{a}) and with the control operation $A_{\mathrm{ctr}}$ (\text{b}) close to the APT in $d=2$ dimensional system. Panel (\text{c}) shows $S_X$ for $d=3$ with  $A_{\mathrm{ctrl}}$. Darker (lighter) tones correspond to $L=60$ ($L=40$) for $d=2$ and $L=24$ ($L=16$) for $d=3$, while the colors denote the measurement rate $p$. For $d=2$, at $p=0.8175 \approx p^{\mathrm{APT}}_c$, we observe power-law decays with exponent $\delta =0.45(1)$ characteristic for the DP class in $d=2$. The bottom insets in (\text{a}), (\text{b}) show collapses of data with exponents consistent with DP class in $d=2$, see Tab.~\ref{Tab1}.
    The upper insets in (\text{a}) and (\text{b}) present the value of entanglement entropy $S_X$ at time $t=4L$ for $p<p_c^\mathrm{APT}$ as a function of the size of the subsystem boundary $\partial X$: (\text{a}) indicates an area-law $S_X\propto |\partial X|$, while (\text{b})  exhibits a volume-law behavior: $S_X\propto  |X|$. 
    For $d=3$ (\text{c}) we find a power law behavior with $\delta=0.65(4)$ around the critical point $p_c^\mathrm{APT}=0.912(1)$. In the inset, we reveal the emerging volume-law scaling, $S_X\propto  |X|$. 
    The results are averaged over more than $10^3$ circuit realizations. 
    }
    \label{fig2ent}
\end{figure*}

\subsection{Entanglement evolution}
\label{sec:4}

Now, we switch to the full quantum dynamics of the circuit and calculate the time-evolved state $\ket{\Psi_t}$. We focus on non-linear functions of the density matrix $\rho_t$, which grasp physics beyond the average state properties. We consider the average entanglement entropy $S_X(t)$ for subsystem $X$ which is a hyper rectangle 
of dimensions $l_x \times L/2 \times \ldots \times L/2$ (recall that the full system has dimensions  $L \times L/2 \times \ldots \times L/2$), and set the value of $l_x$ as $L/4$, which allows us to distinguish between area-law and volume-law scaling of entanglement entropy when the system size $L$ is increased.

Without the feedback-control operation $A_\mathrm{ctrl}$, i.e., in the presence only of the short-range feedback control, the system undergoes an MIT at $p^\mathrm{MIT}_c=0.255(3)$ between phases with volume-law and area-law entanglement entropy, with properties fully analogous to the MIT reported in $d=2$ systems without feedback~\cite{sierant2022measurementinducedphase}. 

In Fig.~\ref{fig2ent}(a), we present time evolution of the entanglement entropy $S_X(t)$ at measurement rates $p$ close to the APT which occurs at $p_c^\mathrm{APT}=0.8175(2)$. Entanglement entropy $S_X(t)$ saturates at $p<p^\mathrm{APT}_c$ to a finite value, decays exponentially with time $t$ when $p>p^\mathrm{APT}_c$ and follows a power-law decay when $p\simeq p^\mathrm{APT}_c$. This behavior is analogous to the defect density $n_\mathrm{def}$ near APT. Moreover, as shown in the lower inset of Fig.~\ref{fig2ent}(a), the entanglement entropy admits the same dynamical scaling as $n_\mathrm{def}$, with compatible critical exponents, see Tab.~\ref{Tab1}. Importantly, the entanglement entropy presents an area-law behavior and scales proportionally to the number of sites at the boundary $\partial X$ of the region $X$, $S_X\propto |\partial X|$ at any measurement probability $p > p_c^\mathrm{APT}$, as illustrated in the upper inset in Fig.~\ref{fig2ent}(a). Consequently, at $p=p^\mathrm{APT}_c$, there is an area-to-area law entanglement transition at times $t\propto L$, in full analogy with the $d=1$ case~\cite{sierantcontrol}.

In the presence of the feedback-control operation $A_\mathrm{ctrl}$, the dynamical behavior of the entanglement entropy $S_X(t)$ at a fixed subsystem size is entirely analogous to the short-range feedback case, as shown in Fig.~\ref{fig2ent}(b). However, in the presence of $A_\mathrm{ctrl}$, entanglement entropy has a volume-law scaling with subsystem size at all $p < p^\mathrm{APT}_c$. This is demonstrated in the upper inset in Fig.~\ref{fig2ent}(b), which shows that $S_X/|X|$ approaches a constant with increasing subsystem size. Thus, at measurement rate $p=p^\mathrm{APT}_c$, the system undergoes an entanglement transition between volume-law scaling and area-law scaling of entanglement entropy at times $t \propto L$ (see the next Section for further discussion of this point). 

Finally, the results for $d=3$, presented in Fig.~\ref{fig2ent}(c) exhibit an analogous behavior. We note that the power-law decay of $S_X/|X|$ close to APT is governed by an exponent $\delta\approx 0.65(4)$, slightly smaller than the value for the DP class for $d=3$. This is a finite-size effect caused by the limitations in the largest system size, $L=24$, available to our simulations of full quantum dynamics. In contrast, the calculations of the average dynamics were performed for systems of size up to $L=128$ at $d=3$ and yielded the result consistent with DP class in $d=3$. The presence of the feedback-control operation $A_\mathrm{ctrl}$ ensures that the entanglement entropy follows a volume-law $S_X \propto |X|$ as indicated by the saturation of the curves shown in the inset of Fig.~\ref{fig2ent}(c). Consequently, at times $t \propto L$, the system undergoes an entanglement transition between volume-law and area-law scaling of entanglement entropy.

The critical features of entanglement entropy dynamics reported in this Section for systems in $d=2$ and $d=3$ dimensions are entirely analogous to results for $d=1$ discussed in~\cite{sierantcontrol}.
By analogy, we expect similar results to extend to $d=4$, the upper critical dimension for the DP universality class~\cite{henkel2008non}. Nevertheless, our present capabilities of the simulation of Clifford circuits prevent us from a quantitative confirmation of this conjecture.

\begin{figure}
    \centering
    \includegraphics[width=0.9\columnwidth]{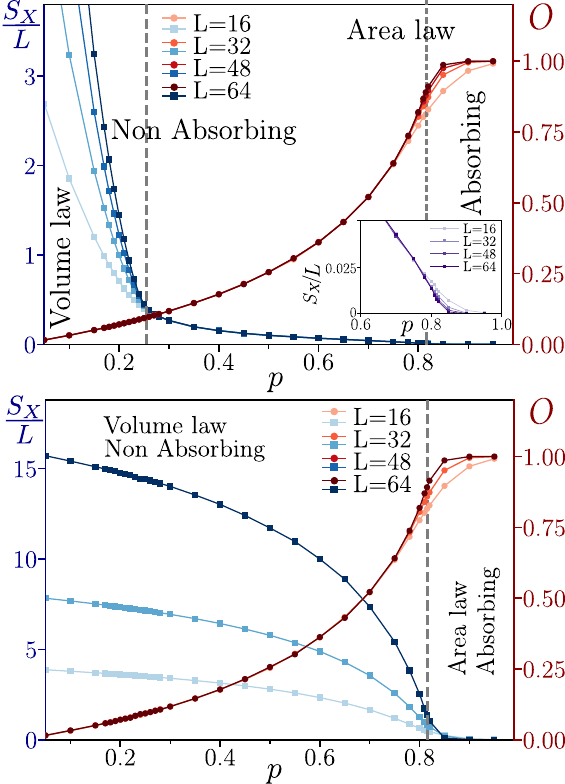}
    \caption{Properties of the system at time $t=4L$ as function of the measurement rate $p$ for $d=2$ dimensional system. The top panel presents results with short-range feedback only (i.e., without the control $A_\mathrm{ctrl}$) and demonstrates a volume-law phase of entanglement $S_A \propto L^2$ at $p < p^{\mathrm{MIT}}_c = 0.255(3)$ and an area-law phase  $S_A \propto L$ at $p > p^{\mathrm{MIT}}_c$. The order parameter $O\equiv 1- n_{\mathrm{def}}$ is smaller than unity for $p < p^{\mathrm{APT}}_c = 0.8165(19)$ and approaches $1$ in the $L \to \infty$ limit for $p > p^{\mathrm{APT}}_c$. The inset in the top panel shows the area-to-area law transition of the entanglement entropy $S_X$ at  $p=p^{\mathrm{APT}}_c$.
    The bottom panel -- the same, but in the presence of the global feedback-control operation $A_{\mathrm{ctrl}}$. The APT and MIT merge in a single transition at which the volume-to-area law entanglement transition accompanies the ordering transition reflected by $O$. 
    \label{fig:3}
}
\end{figure}
\subsection{Absorbing and entanglement phase transition at time $t\propto L$} 
\label{sec:5}

As we argued in Sec.~\ref{subsec:phenom},  the presence of the feedback mechanism in our system implies that $\lim_{t\to\infty}{\rho_t} = |\mathrm{ABS}\rangle\langle\mathrm{ABS}|$ at any finite system size $L$. In other words, if the limit of large time is taken, we will always find our system in the trivial, ordered, product state $|\mathrm{ABS}\rangle\langle\mathrm{ABS}|$.
However, fixing a specific time scale at which we observe the system, for instance, setting $t=4L$ (which we will use henceforth in this Section), allows us to uncover manifestations of the dynamical phase transitions described above.

In Fig.~\ref{fig:3}, we compare the results for $d=2$, considering both short-range feedback and including the global feedback-control operation $A_\mathrm{ctrl}$. A clear signature of the APT  is the fact that the order parameter $O \equiv 1- n_\mathrm{def}$ approaches its maximal value $O\to 1$ at $p>p^\mathrm{APT}_c$ with increasing system size, while $O<1$ for $p<p_c^\mathrm{APT}$, as demonstrated by the red lines in Fig.~\ref{fig:3}. We reiterate that the behavior of the order parameter is the same in the presence and absence of $A_\mathrm{ctrl}$ by the construction of our feedback mechanism. 

In presence of the short-range feedback only, see the top panel in Fig.~\ref{fig:3}, we observe two separate transitions: the MIT between phases with volume-law and area-law entanglement entropy at measurement rate $p=p^{\mathrm{MIT}}_c = 0.255(3)$ and the APT at $p=p^{\mathrm{APT}}_c$. We note that area-law scaling for $d=2$ implies that $S_X\propto L$. Notably, at $p=p^{\mathrm{APT}}_c$, there is a second entanglement transition between the area-law phase characterized by $S_X\propto L$ and an area-law phase with $S_X \to 0$ as demonstrated by the inset in the top panel of Fig.~\ref{fig:3}.

In contrast, when the global feedback-control operation $A_\mathrm{ctrl}$ is present, the entanglement entropy $S_X$ behavior parallels that of the defect number $n_\mathrm{def}$. Consequently, in that case, there is only a single volume-law to area-law entanglement transition in our system at $p=p^{\mathrm{APT}}_c$, as shown in the lower panel of Fig.~\ref{fig:3}. 

The behavior of entanglement entropy presented at times $t=4L$ has its roots in the separation of time scales of the approach to the absorbing state that occurs at the APT. Since the dynamical behavior of entanglement entropy in $d=3$ parallels the results for $d=2$, the entanglement entropy has analogous behavior to the one presented in Fig.~\ref{fig:3} in the presence and in the absence of the global feedback control operation in $d=3$ at times $t\propto L$ (data not shown).

\begin{table}
    \begin{tabular}{c|cc|cc|cc}
        \hline \hline
         & \multicolumn{2}{c}{DP class} &   \multicolumn{2}{c}{Unitary Dyn.} & \multicolumn{2}{c}{Average Dyn.} \\
        \cline{2-3} \cline{4-7}
        {$d$} & {$\delta$} & {$\nu$} & {$\delta$} & {$\nu$} &  {$\delta$} & {$\nu $}\\
        \hline
        2 & 0.450(5) & 1.295(6) & 0.45(2) & 1.30(5) & 0.45(1) & 1.30(3)  \\
        3 & 0.73(1) &  1.11(1) & 0.65(4)$^*$ &  -  & 0.73(2)  & 1.09(4) \\ 
        4 & 1 & 1 & - & -  & 0.85(5)$^*$  & 1.0(1)$^*$  \\ 
        \hline \hline
    \end{tabular}
    \caption{Summary of the critical exponents characterizing dynamical transitions in flagged Clifford circuits. The column denoted by "DP class" shows the expectations of directed percolation theory~\cite{Munoz99, Hinrichsen00}. The column "Unitary Dyn." presents the exponents obtained in studies of the full quantum dynamics of the flagged stabilizer circuits. At the same time, the column "Average Dyn." reports the results obtained by simulation of the probabilistic cellular automaton that corresponds to the average dynamics of the flagged stabilizer circuits.
    The results of our numerical simulations are consistent with the DP universality class except for the average dynamics results for $d=4$, which are subject to significant system size drifts and hence are denoted by the asterisks.
    }
    \label{Tab1}
\end{table}

\section{Conclusion}
\label{sec:6}

In this work, we analyzed the role of dimensionality in the dynamics of monitored stabilizer circuits with a feedback control mechanism introduced by classical labels (flags), which gives rise to an absorbing state of the dynamics. While dimensionality changes the universal content of APT and MIT, the phenomenological understanding presented for one-dimensional circuits is generalized straightforwardly to higher dimensions $d$. In particular, similarly to $d=1$ dimensional case~\cite{sierantcontrol}, the range of feedback-control operations is a crucial ingredient for the interplay between entanglement and absorbing state transition. 
We find that circuits with short-range feedback control exhibit two entanglement transitions at circuit depths proportional to the system size: a volume-to-area law transitions at the MIT critical point and an area-to-area law transition at the APT transition point. 
Instead, when global feedback-control operations are present, there is only a single volume-to-area MIT which coincides with the APT. For the employed global unitary operation, MIT inherits the properties of the underlying APT universality class. 
In our implementation, the latter is unaffected by the feedback control operations range and always leads to critical behavior described by the directed percolation universality class in $d$ dimensions, as summarized in Tab.~\ref{Tab1}. The average dynamics results are consistent with the expectation that $d=4$ is the upper critical dimension beyond which mean-field critical exponents capture properties of APT. 
In contrast, the upper critical dimension for MIT in setups without feedback mechanisms is $d_c=6$~\cite{sierant2022measurementinducedphase}. 

Similarly to the one-dimensional case, our work concludes that the post-selection problem can be mitigated if appropriate feedback-control operations are chosen. The behavior of the entanglement entropy at the area-to-area law phase transition in the setup with short-range feedback control and the volume-to-area law phase transition in the setup with global control operation can be observed by measurements of the defect density, which does not require the post-selection. However, the crucial caveat is that the correspondence between the dynamics defect density and the entanglement entropy does not generally hold but requires a choice of sufficiently strongly entangling control operation (see~\cite{sierantcontrol} for explicit examples).
In other words, the post-selection problem is mitigated only by meeting stringent control operations requirements. 
Moreover, introducing global control operations may significantly alter the trajectory ensemble. As a result, the feedback control drives the original measurement-induced transition (present in the system without feedback control) onto a different universality class. 
We expect similar conclusions to hold for generic (Haar) circuits. While numerical methods are ineffective, generalizing the arguments in Ref.~\cite{piroli} may lead to a formal proof of the distinct APT and MIT when $ q\gg1 $ dimensional qudits are considered. 
Similarly, we expect that our arguments generalize to the monitored fermionic model with conditional feedback control in higher dimensions and variable range interactions, as we will extensively discuss in a future contribution~\cite{sierant2024}.
An interesting future direction is to enhance our understanding of the interplay between absorbing states, topological state preparation, and shallow circuits~\cite{lavasani2021measurementinducedtopological,lavasani2021topologicalorderand,zhu2022nishimoriscat,sang2021measurementprotectedquantum,lee2022decodingmeasurementpreparedquantum,klocke2022topologicalorderand,zhu2023structured,klocke2023majorana}. We leave these questions as subjects of further explorations. 

\begin{acknowledgments}
XT acknowledges support from the ANR grant "NonEQuMat."
(ANR-19-CE47-0001) and computational resources on the Coll\'ege de France IPH cluster. 
PS acknowledges support from: ERC AdG NOQIA, Ministerio de Ciencia y
Innovation Agencia Estatal de Investigaciones (PGC2018-097027-B-
I00/10.13039/501100011033,  CEX2019-000910-S/10.13039/501100011033, Plan National
FIDEUA PID2019-106901GB-I00, FPI, QUANTERA MAQS PCI2019-111828-2,
QUANTERA DYNAMITE PCI2022-132919, Proyectos de I+D+I “Retos Colaboración”
QUSPIN RTC2019-007196-7); MICIIN with funding from European Union
NextGenerationEU(PRTR-C17.I1) and by Generalitat de Catalunya; Fundació Cellex;
Fundació Mir-Puig; Generalitat de Catalunya (European Social Fund FEDER and CERCA
program, AGAUR Grant No. 2017 SGR 134, QuantumCAT \ U16-011424, co-funded by
ERDF Operational Program of Catalonia 2014-2020); EU Horizon 2020 FET-OPEN OPTOlogic (Grant No 899794); EU Horizon Europe Program (Grant Agreement 101080086 — NeQST), National Science Centre, Poland (Symfonia Grant No. 2016/20/W/ST4/00314); ICFO Internal
"QuantumGaudi" project; European Union's Horizon 2020 research and innovation program
under the Marie-Skłodowska-Curie grant agreement No 101029393 (STREDCH) and No
847648 (“La Caixa” Junior Leaders fellowships ID100010434: LCF/BQ/PI19/11690013,
LCF/BQ/PI20/11760031, LCF/BQ/PR20/11770012, LCF/BQ/PR21/11840013). 

Views and
opinions expressed in this work are, however, those of the author(s) only and do not
necessarily reflect those of the European Union, European Climate, Infrastructure and
Environment Executive Agency (CINEA), nor any other granting authority. Neither the
European Union nor any granting authority can be held responsible for them.
\end{acknowledgments}

\appendix
\section{Additional details}
For self-consistency, this Appendix presents additional technical details. 
After a more formal discussion on flagged Clifford circuits, we discuss mapping the average dynamics to a classical model. We briefly review the Gottesman-Knill theorem and how stabilizer simulations are performed. 

\onecolumngrid
\subsection{Flagged Clifford circuits}
Our discussion follows closely Ref.~\cite{sierantcontrol}, to which we refer for additional details. 
At a formal level, flagged Clifford circuits extend our many-body quantum state to $|\Psi\rangle\mapsto |\Phi\rangle = |\Psi\rangle\otimes |\mathcal{F}\rangle$. The flag vector $|\mathcal{F}\rangle = \otimes_{\vec{m}\in \Lambda}|f_{\vec{m}}\rangle$ registers the post-measurement polarizations and determines the action of the unitary gates at each time step. 
As for the Main Text, we fix $|\mathrm{ABS}\rangle=|1_{\vec{m}}\rangle$ as the absorbing state. 

The key idea is that flagged sites $\vec{m}$ (i.e., for which $f_{\vec{m}}=1$) are unchanged by the measurement layer $M$ and by the global unitary $A$. Furthermore, depending on the nearest neighbors flags $f_{\vec{n}}$, the two-body gates $U_{\vec{m},\vec{n}}$ act trivially or as a random Clifford transformation. (Here $\vec{n} = \vec{m}+\hat{\vec{e}}_\mu$, with  $\hat{\vec{e}}_\mu$ being the versor in the randomly chosen direction $\mu=1,\dots,d$).

In the doubled Hilbert space, the absorbing state $|\Phi_{\mathrm{abs}}\rangle = |\mathrm{ABS}\rangle\otimes|1_{\vec{m}}\rangle^{\otimes_\vec{m}}$ is the fixed point of the dynamics, while the initial state is $|\Phi_0\rangle = |\Psi_0\rangle\otimes|0_{\vec{m}}\rangle^{\otimes_\vec{m}} $. 
The non-trivial control operation $A_\mathrm{ctrl}$, when present, is given by 
\begin{equation}
    A_\mathrm{ctrl} |\Phi\rangle = (C_{\{\vec{m}:f_{\vec{m}}=0\}} |\Psi\rangle )\otimes |\mathcal{F}\rangle 
\end{equation}
with $C_{\{\vec{m}:f_{\vec{m}}=0\}}$ a global Clifford unitary acting only on unflagged sites. 
On the other hand, the projective measurement and two-body gates are respectively given by 
\begin{equation}
    P_{\vec{m}} |\Phi\rangle = \begin{cases}
        \displaystyle\frac{1}{\sqrt{p_-}} \left(\frac{1-Z_{\vec{m}}}{2}|\Psi\rangle\right)\otimes \left[(X_{\vec{m}}-i Y_{\vec{m}})^{f_{\vec{m}}}|\mathcal{F}\rangle \right], &\text{outcome }-1,\\
        \displaystyle\frac{1}{\sqrt{p_+}} \left(\frac{1+Z_{\vec{m}}}{2}|\Psi\rangle\right)\otimes \left[(X_{\vec{m}}+i Y_{\vec{m}})^{1-f_{\vec{m}}}|\mathcal{F}\rangle \right], &\text{otherwise},
    \end{cases}
\end{equation}
and for nearest neighboring sites $\vec{m}$ and $\vec{n}$
\begin{equation}
    U_{\vec{m},\vec{n}}|\Phi\rangle  = \begin{cases}
       C_{{\vec{m},\vec{n}}} |\Psi\rangle\otimes  \left[(X_{\vec{m}}-i Y_{\vec{m}})^{f_{\vec{m}}}(X_{\vec{n}}-i Y_{\vec{n}})^{f_{\vec{n}}}|\mathcal{F}\rangle \right], &\text{if } f_{\vec{m}}f_{\vec{n}}=0 \\
       |\Phi\rangle , &\text{otherwise}.
    \end{cases}
\end{equation}

\subsection{Mapping of the average dynamics to a classical model}
\label{app:mapping}
As discussed in the previous Section, the flagged Clifford circuit acts in a formally doubled Hilbert space. We now discuss the dynamics of the average state over Clifford gates. 
Crucially, the average dynamics of observables linear in the state such as $n_\mathrm{def}$ is fully encoded in the mean state $R_t = \mathbb{E}_{\mathrm{Clifford}}[|\Phi_t\rangle\langle\Phi_t|]$, with $t$ the depth/time of the circuit. 
The core idea is that the average dynamics over the Clifford unitaries correspond to a probabilistic cellular automaton (PCA). The average state requires independently drawn Clifford operations $C$, each of them acting on a single bra and ket, namely
\begin{equation}
    I \equiv  \int_{\mathrm{Clifford}} dC C |\Psi\rangle\langle \Psi| C^\dagger.
\end{equation}
However, the integral $I$ is easily performed using the 2-design property of the Clifford group~\cite{Gross2021} as the Haar integral
\begin{equation}
    I = \int_{\mathrm{Haar}} dC C |\Psi\rangle\langle \Psi| C^\dagger = \frac{1}{2^w} \openone_{C}\otimes  P^\perp |\Psi\rangle\langle \Psi|P^\perp,
\end{equation}
where $\openone_C$ is the sites where $C$ act non-trivially, and $P^\perp$ is the projection on its complementary space. 
Introducing the indices $k_{\vec{m}} = 0,\pm 1$ the on-site mixed state $\rho_{\vec{m}}^{(0)} = \openone_{\vec{m}}/2$ and the on-site projectors $\rho_{\vec{m}}^{(\pm 1 )} = (\openone_{\vec{m}}\pm Z_{\vec{m}})/2$, it follows that the Clifford averaged state is
\begin{equation}
    R_t = \mathbb{E}_{\mathrm{Clifford}}[|\Phi_t\rangle\langle\Phi_t|] = \left(\bigotimes_{\vec{m}\in \Lambda} \rho_{\vec{m}}^{(f_{\vec{m}})} \right)\otimes \left(\bigotimes_{\vec{m}\in \Lambda} \rho_{\vec{m}}^{(-1)^{f_{\vec{m}}+1}}\right).
\end{equation}
Since the physical state and the flags are in one-to-one correspondence, the average dynamics correspond to a probabilistic cellular automaton of the flags. 
The average dynamics over measurement locations, measurement outcomes, and unitary locations can be written down analogously. It corresponds to a discrete master equation for $R_t$, that we do not detail for presentation purposes, cf. also~\cite{sierantcontrol}. 

\subsection{Brief review on the simulation of stabilizer circuits}
We conclude by briefly reviewing the ideas for stabilizer simulations and refer to~\cite{aaronson2004improvedsimulationof,sierant2022measurementinducedphase} for additional details. 
A stabilizer state on the lattice $\Lambda$ is fixed by $L^d$ independent Pauli strings $g_{\vec{m}}$ such that $g_{\vec{m}}|\Psi\rangle = |\Psi\rangle$. Each Pauli string is parametrizes as 
\begin{equation}
    g_{\vec{m}} = e^{i \pi \phi_{\vec{m}}} \prod_{\vec{j}\in \Lambda}(X^{a_{\vec{m}}^{{\vec{i}}}} Z^{b_{\vec{m}}^{{\vec{i}}}})
\end{equation}
where $\phi_{\vec{m}}, a_{\vec{m}}^{{\vec{i}}}, b_{\vec{m}}^{{\vec{i}}} = 0,1$ are $\mathbb{Z}_2$ numbers. 
The group $\mathcal{G}$ generated by the Pauli strings $g_{\vec{m}}$ is abelian and fixes the state as $|\Psi\rangle\langle \Psi| = \sum_{g\in \mathcal{G}} g / 2^{L^d}$. Therefore, the state is completely determined by the matrix $G = (\phi_{\vec{m}}| a_{\vec{m}}^{{\vec{i}}}, b_{\vec{m}}^{{\vec{i}}})$, whose rows fix the generators of the group $\mathcal{G}$. 

Stabilizer circuits involve stabilizer states that evolve under Clifford gates and projective measurements. 
By definition, Clifford unitaries transform a Pauli string in a single Pauli string. Hence, they correspond to a transformation of the $L^d\times (L^d+1)$ matrix $G$ to a new matrix $G'$~\cite{aaronson2004improvedsimulationof}. 
Similarly, projective measurements onto Pauli strings transform $G$ in a new matrix $G''$. If the projecting Pauli string $g^*$ is already in the group $\mathcal{G}$, then $G\mapsto G$. (Finding the measurement outcome requires a Gaussian elimination, cf.~\cite{aaronson2004improvedsimulationof}). Viceversa, if the operator $g^*$ is not in the group $\mathcal{G}$ then there exist a set $ I_\mathrm{anti}$ of anticommuting operators $g_\mu$ such that $\{  g^*,g_\mu\} = 0$  for each $\mu\in I_\mathrm{anti}$. The measurement outcome is randomly and uniformly $\pm 1$, and the state collapses after the measurement onto the resulting string $\pm g^*$. One can verify that the updated matrix $G\mapsto G''$ is given by $g_\nu$ such that $[g^*,g_\nu]=0$, together with $g^*$ and the transformed set ${g}_{\tilde{\mu}}\cdot g_\mu$ for $\tilde{\mu} \in I_{\mathrm{anti}}$ and $\mu \in I_{\mathrm{anti}}/\{\tilde{\mu}\}$. 
The above statement summarizes the Gottesman-Knill theorem and illustrates how the system is efficiently simulable with polynomial classical resources in the number of qubits $N$. 
Lastly, given a bipartition $X\cup X_c$ the entanglement of a stabilizer state $|\Psi\rangle$ can be computed efficiently via~\cite{hamma2004,PhysRevX.7.031016} as
\begin{equation}
    S_X(|\Psi\rangle) = |X| - \log_2|\mathcal{G}_X|,
\end{equation}
with $\mathcal{G}_X$ the subgroup of all elements in $\mathcal{G}$ that act trivially on $X_c$, and $|X|$ is the number of qubits in $X$. The calculation of $\log_2|\mathcal{G}_X|$ reduces to the calculation of the rank of an appropriate submatrix of the matrix $G$ over the $\mathbb{Z}_2$ field for which we use the algorithm of ~\cite{andren2007onthecomplexity,albrecht2011efficient,Bertolazzi14}. We note that participation entropies of stabilizer states can be calculated in a similar fashion, see \cite{sierant2022universalbehaviorbeyond}.

\twocolumngrid

\end{document}